\documentclass[preprint,12pt]{elsarticle}

\usepackage{moreverb,url}
\usepackage[colorlinks,bookmarksopen,bookmarksnumbered,citecolor=red,urlcolor=red]{hyperref}

\usepackage{amsmath}
\usepackage{graphicx}
\usepackage{multicol}
\usepackage{multirow}
\usepackage{array}
\usepackage{floatrow}
\usepackage{subfigure}
\usepackage{url}
\begin{document}

\title{A Modified Global Climate Simulation Model}

 \author[1]{Sofya Prakhova$^{\ast}$}
 \author[1]{Volker Rehbock} 
 \author[2]{Igor Suleimanov}
 
 \address[1]{Department of Mathematics and Statistics, Curtin University, GPO Box U1987, Perth, Australia, 6845}
 \address[2]{Department of Mathematics, Ufa State Petroleum Technological University, 1 Kosmonavtov street, Ufa, Russia, 450062}
 \cortext[cor1]{Corresponding author.
 	\textit{Email address:} {sofya.prakhova@gmail.com}}

%
%

%
%
%
%

\begin{abstract}
In this paper, we incorporate seasonal variations of insolation into the global climate model C-GOLDSTEIN. We use a new approach for modelling insolation from the space perspective presented in the authors\textquoteright \space earlier work and build it into the existing climate model. 

Realistic monthly temperature distributions have been obtained after running C-GOLDSTEIN with the new insolation component. Also, the average accuracy of modelling the insolation within the model has been increased by 2\%.  In addition, new types of experiments can now be performed with C-GOLDSTEIN, such as the investigation of consequences of random variations of insolation on temperature etc. 
\end{abstract}


\maketitle

\section{Introduction}
The development of climate models started back in the early 1960s, when the first models containing only the atmosphere appeared \cite{manabe1965simulated}. But as time was progressing and the computational capacities increased, more components were developed and coupled together. The resulting comprehensive models are known as General Circulation Models (GCMs). These models represent powerful tools for predicting future climate changes, as well as for understanding the climate of the past. The examples of the recent GCMs are MIROC-ESM-CHEM \cite{watanabe2010improved}, CNRM-CM5 \cite{voldoire2013cnrm}, IPSL \cite{marti2010key}, CSM4 \cite{gent2011community}, CMCC-CESM \cite{vichi2011global}.
Parallel to them, another group of models was developing- the Earth System Models of Intermediate Complexity (EMICs). These models are more simplified than the comprehensive GCMs. However, they have a number of advantages, such as their capability to be used for the forecasts up to several millennia, as well as for performing extensive sensitivity studies. The examples of the EMICs are EcBilt \cite{opsteegh1998ecbilt}, IGSM2 \cite{sokolov2005integrated}, CLIMBER-2 \cite{petoukhov2000climber}, DCESS \cite{shaffer2008presentation}, Bern3D-LPJ \cite{ritz2011coupled}, LOVECLIM1.2 \cite{goosse2010description}.

In this paper, we incorporate the seasonal variations of insolation into the EMIC C-GOLDSTEIN \cite{marsh2002development}, which previously used yearly averages of insolation. In order to do this, the annual averages of insolation have been replaced by approximation curves of insolation at any particular time. The approximation was done by using the least square method based on the results obtained from the authors\textquoteright \space earlier work \cite{prakhova2014new}, where a new approach for modelling insolation has been proposed. Realistic monthly latitudinal temperature distributions have been obtained after running the C-GOLDSTEIN model with the new insolation component. The average accuracy of modelling the insolation within the model has been increased from 96\% to 98\%. In addition, this work broadens the applications of C-GOLDSTEIN, because calculations can now be performed for any particular time of the year. 

\section{Description of the model}

C-GOLDSTEIN (Global Ocean-Linear Drag Salt and Temperature Equation INtegrator) consists of a two-dimensional atmospheric model, a three-dimensional ocean model, and simple land surface and sea ice models. The full description of the model is provided in Marsh et al. \cite{marsh2002development}. Longitudinal resolution of the atmospheric component is $10^\circ$, while latitudinal resolution varies from $3^\circ$ near the equator to $20^\circ$ for polar regions. 

The ocean component is based on thermocline equations with an additional linear drag term in the horizontal momentum equations. A condition of zero normal fluxes of heat and salt was specified at the lateral boundaries. The lower boundary fluxes of two prognostic variables (temperature and salinity) were set to zero. 

The land component has no dynamical land-surface scheme and only determines the runoff of fresh water. The surface temperature was assumed to be equal to the atmospheric temperature and the evaporation is set to zero. The sea ice component contains dynamic equations which were solved for the fraction of the ocean surface covered by sea ice and the average height of sea ice. 

The atmospheric component of the model is represented by an Energy Moisture Balance Model. The prognostic parameters are air temperature and specific humidity at the surface. The model balances heat and moisture within the atmosphere. The net flux of longwave radiation into the atmosphere was modelled as a function of the surface and atmospheric emissivities, the temperature of the underlying surface and the Stefan-Bolzmann constant. The incoming radiation was approximation by Legendre polynomials \cite{north1975analytical} and produces latitudinal-dependent annual average values: 

\begin{align}
S(x)\cong1+S_2P_2(x) \notag
\end{align}
where $S(x)$ in the mean annual distribution of radiation reaching the top of the atmosphere, x is the sine of latitude, $S_2=-0.477$ is a constant, and $P_2(x)=\dfrac{1}{2}(3x^2-1)$  is the second Legendre polynomial \cite{north1981energy}.

Within the model both short-term and multi-millennium forecasts can be performed within a relatively short computational time. The standard time step used for calculations is 0.73 days for the atmosphere and double that for the ocean. In order to obtain near present-day climate, a 2000 year experiment needs to be performed (known as SPINUP) which starts from some unrealistic conditions (such as zero mean global air temperature) and then progresses until the system comes close to equilibrium. 

\section{Curve fitting procedure for incorporating the seasonality into the global climate model}\label{Curve}
\begin{figure}
	\begin{center}
		\subfigure[Northern Hemisphere.]{
			\resizebox*{9.3cm}{!}{\includegraphics{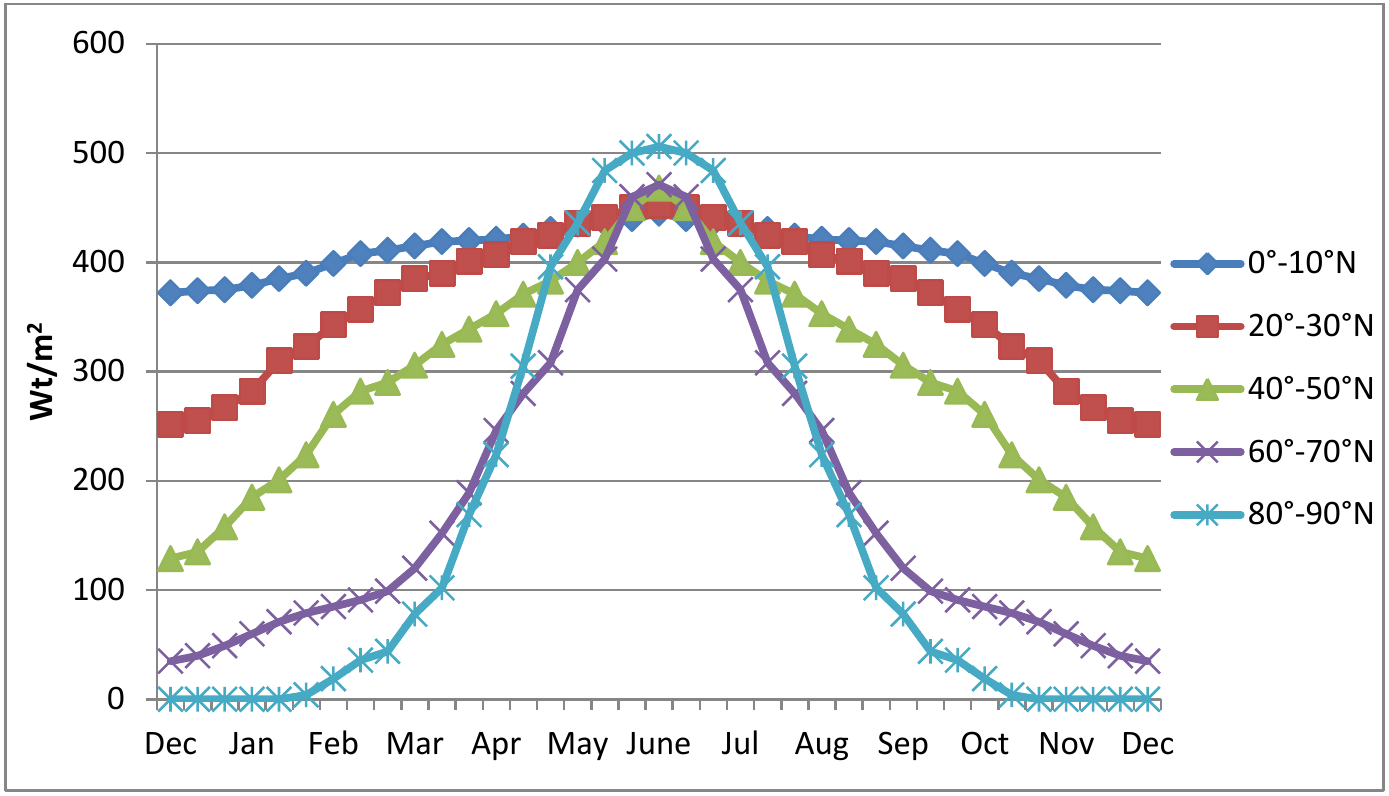}}}	
		\subfigure[Southern Hemisphere.]{
			\resizebox*{9.3cm}{!}{\includegraphics{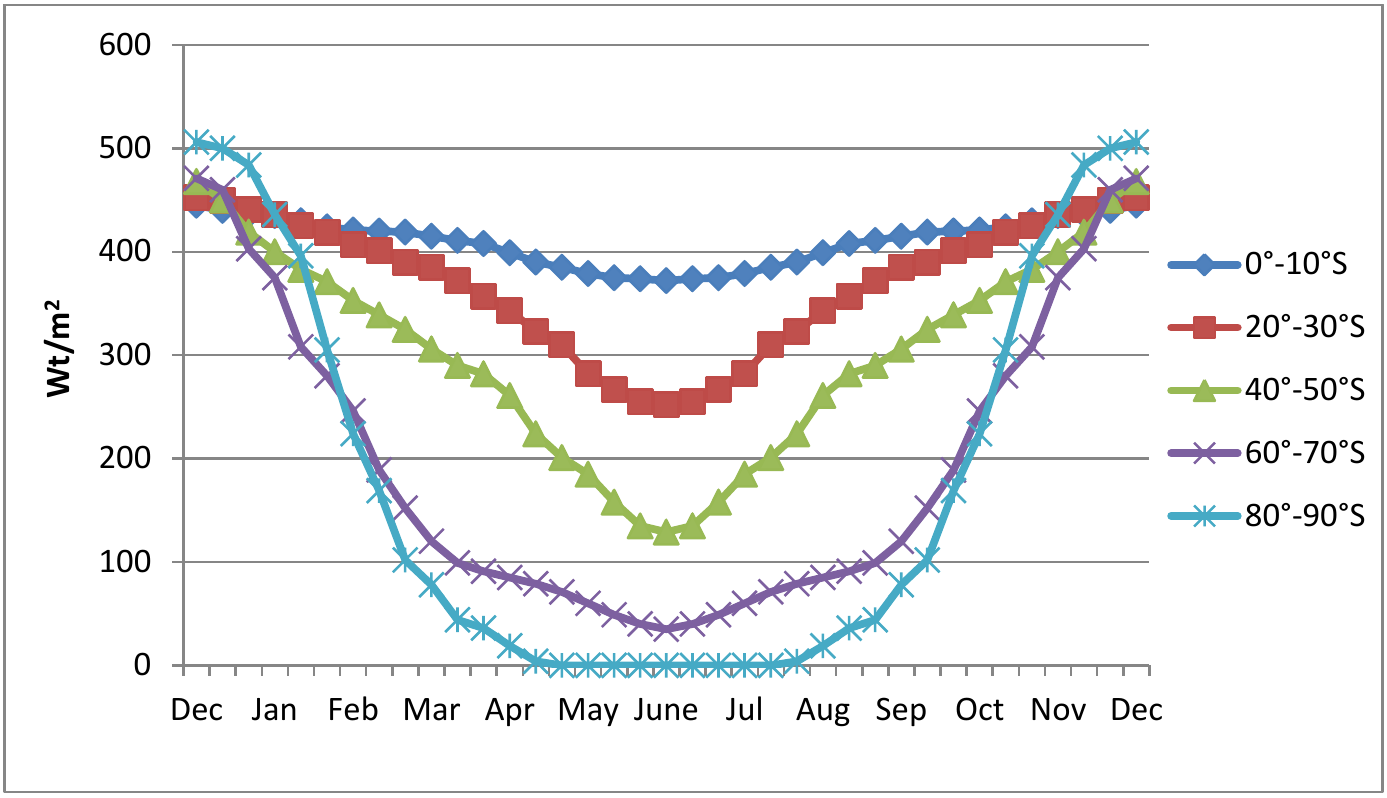}}}
		\caption{The amount of radiation received by odd latitudinal belts in the Northern and Southern Hemisphere.}
		\label{fig:1}
	\end{center}
\end{figure}

In order to incorporate the seasonality into C-GOLDSTEIN model, the amount of insolation for every latitudinal belt throughout the year computed in the authors’ earlier paper \cite{prakhova2014new} was used. The amount of insolation for the odd latitudinal belts of the Northern and Southern Hemisphere is presented in Figure \ref{fig:1}.  

%

In order to allow the incorporation of those curves into the code of the C-GOLDSTEIN, they were approximated by functions of several different types. In particular, the curves corresponding to the first two latitudinal belts ($0^\circ-10^\circ$  and $10^\circ-20^\circ$), which have the least variation, were approximated by a wave function. The curves corresponding to the remaining latitudinal belts were approximated by piecewise continuous functions. In particular, the $30^\circ-40^\circ$, $40^\circ-50^\circ$, $60^\circ-70^\circ$ and $70^\circ-80^\circ$ latitudinal belt curves were best approximated by straight line sections; the best fit for the $20^\circ-30^\circ$ latitudinal belt was a combination of a wave function and the straight lines. The remaining latitudinal belts which displayed a more complicated shape ($50^\circ-60^\circ$ and $80^\circ-90^\circ$) were approximated by the combination of several wave functions and straight lines. 

The wave functions used for the approximation are of the following form: 
\begin{align*} 
\label{eq1} 
y(t)= A\sin(\omega\*t+\varphi)+B 
\end{align*}

The coefficients of the straight lines were found by simply interpolating two given points. In order to find the amplitudes and the vertical shifts of the wave function, the ordinary least square method was used. The optimisation was performed in MS Excel using \textquotedblleft The Solver\textquotedblright \space add-in. The GRG (Generalized Reduced Gradient) non-linear solving method was used.

Note that the Solver command only determines locally optimal solutions and without the specified bounds, a physically unreasonable solution can result. Thus the estimates for the amplitudes and estimates for the vertical shifts were calculated. The estimates for the amplitude were calculated as a half of the difference between the largest and the smallest value of the initial curve on the interval over which the approximation was made. In case of a vertical shift, their sum was taken instead. Based on these, upper bounds for each amplitude and lower bounds for each vertical shift were then chosen so as to allow a reasonable range for the parameters to be optimised.

The angular velocities were fixed as $\omega=\dfrac{2\pi}{p}$, where $p$ denotes the number of intervals over which the wave function is defined. The values for the phases were chosen manually after examining the plot obtained after the first round of optimisation. In case of the optimal values reaching the constraints, the corresponding bounds were shifted further in order to allow an improved and physically reasonable solution to be obtained. 

After performing the optimisation in this way, an optimal solution was reached. For the Southern Hemisphere, a simple shift of the curves for the Northern Hemisphere by 6 months has been made.

\begin{figure}
	\begin{center}
		\subfigure[$0^\circ-10^\circ$N latitudinal belt.]{
			\resizebox*{6.6cm}{!}{\includegraphics{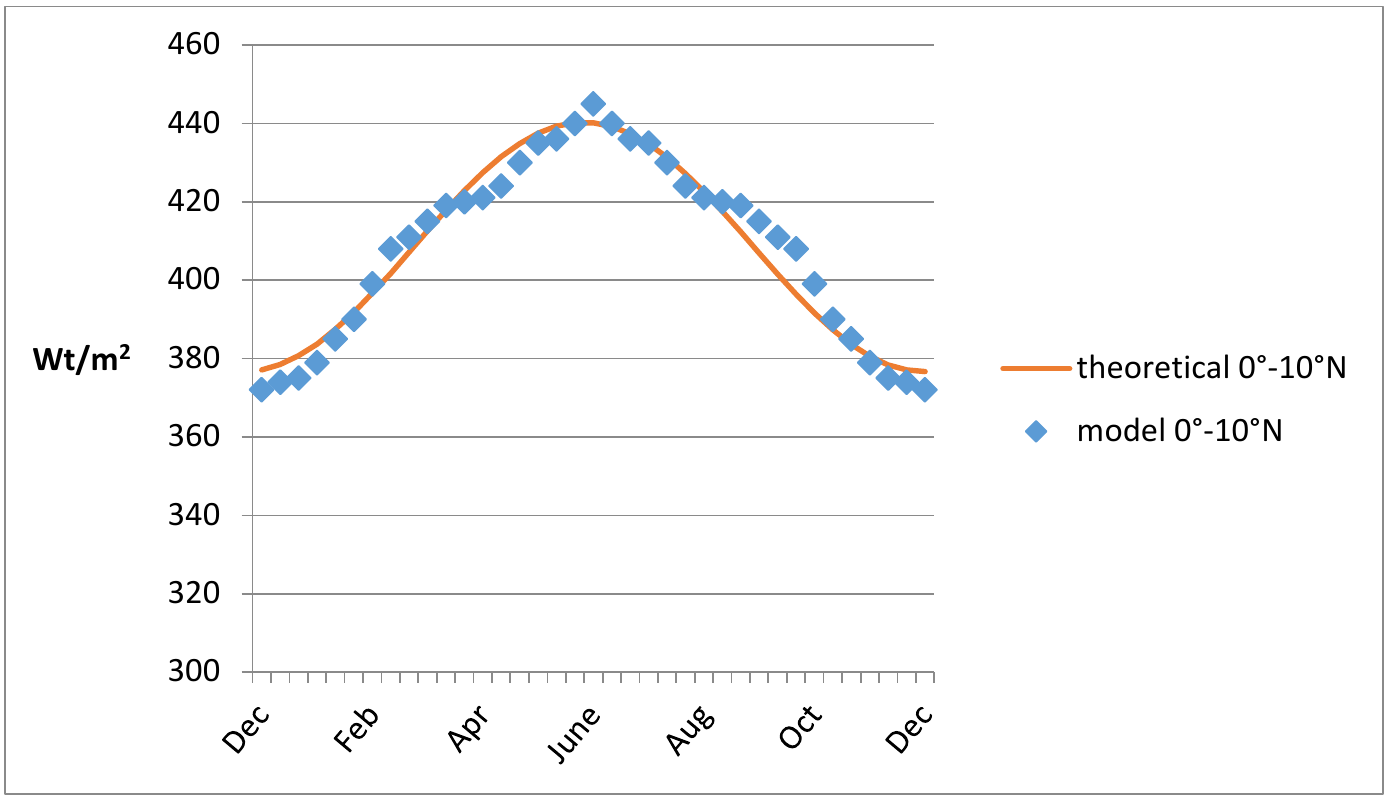}}}	
		\subfigure[$0^\circ-10^\circ$S latitudinal belt.]{
			\resizebox*{6.6cm}{!}{\includegraphics{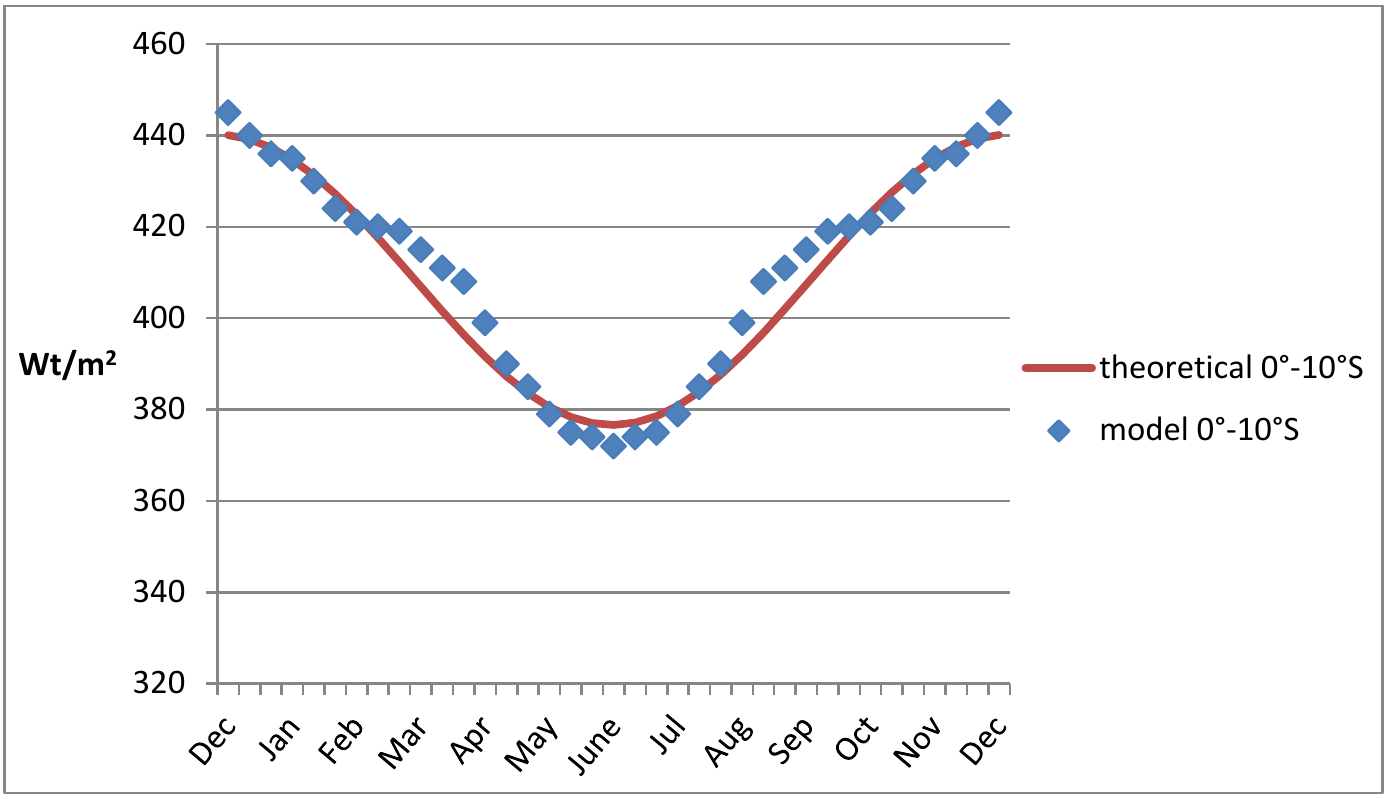}}}
		
		\subfigure[$40^\circ-50^\circ$N latitudinal belt.]{
			\resizebox*{6.6cm}{!}{\includegraphics{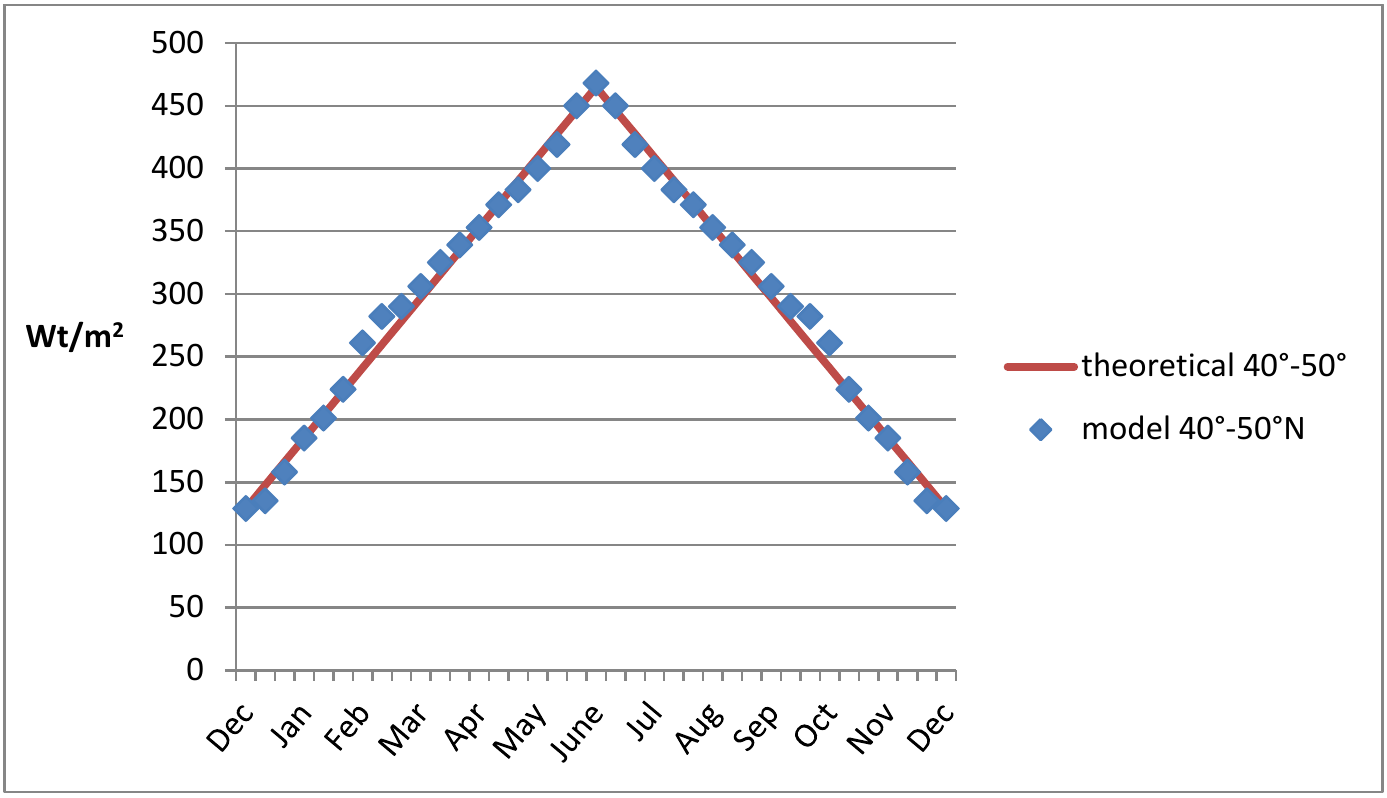}}}	
		\subfigure[$40^\circ-50^\circ$S latitudinal belt.]{
			\resizebox*{6.6cm}{!}{\includegraphics{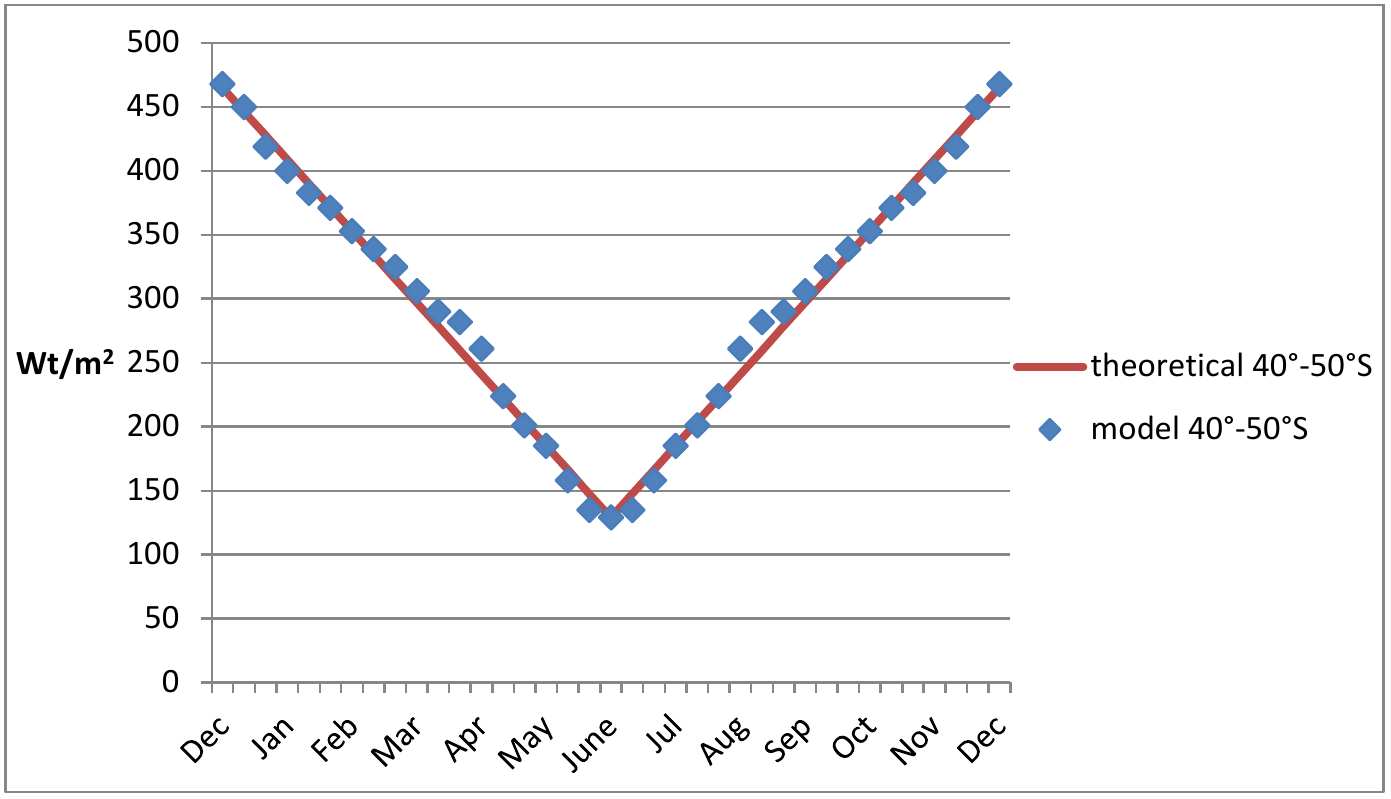}}}
		
		\subfigure[$80^\circ-90^\circ$N latitudinal belt.]{
			\resizebox*{6.6cm}{!}{\includegraphics{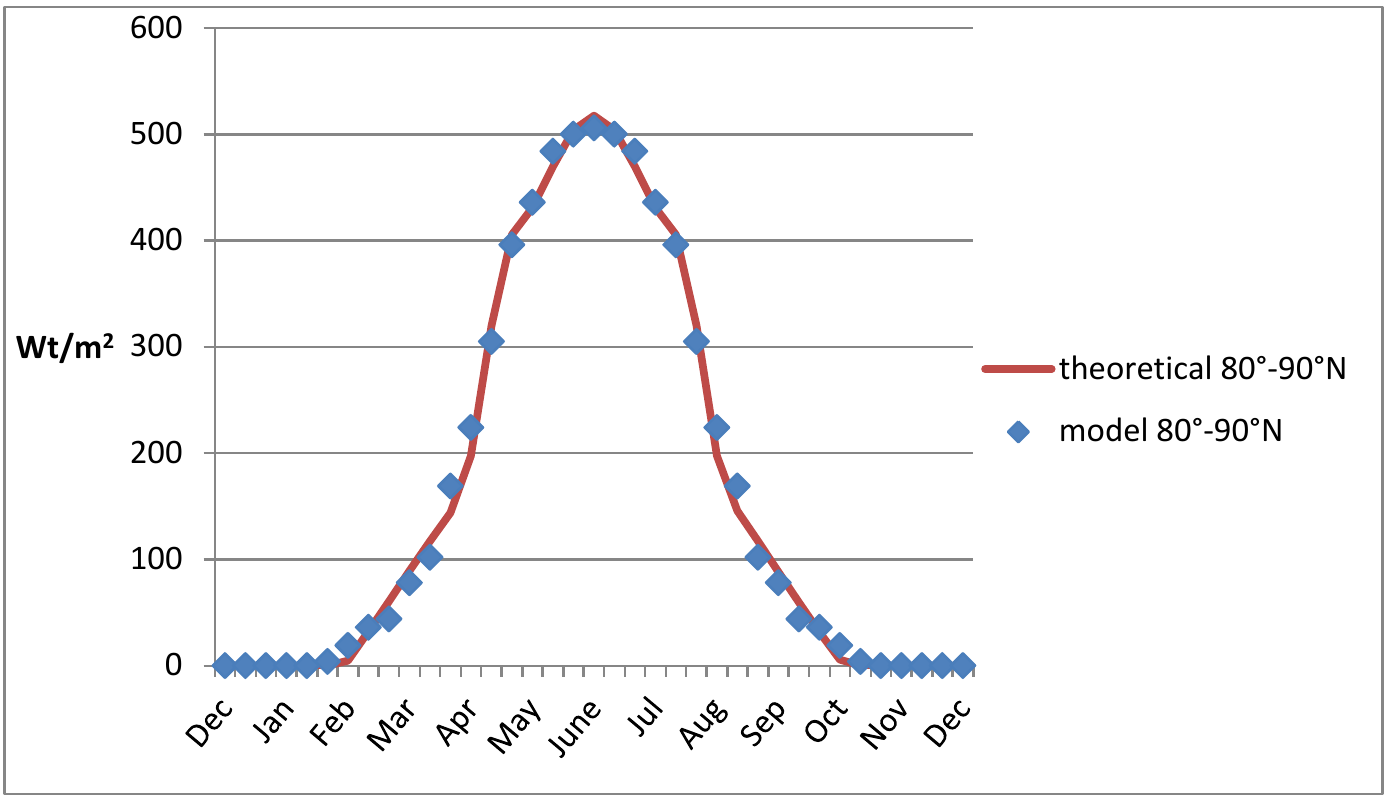}}}	
		\subfigure[$80^\circ-90^\circ$S latitudinal belt.]{
			\resizebox*{6.6cm}{!}{\includegraphics{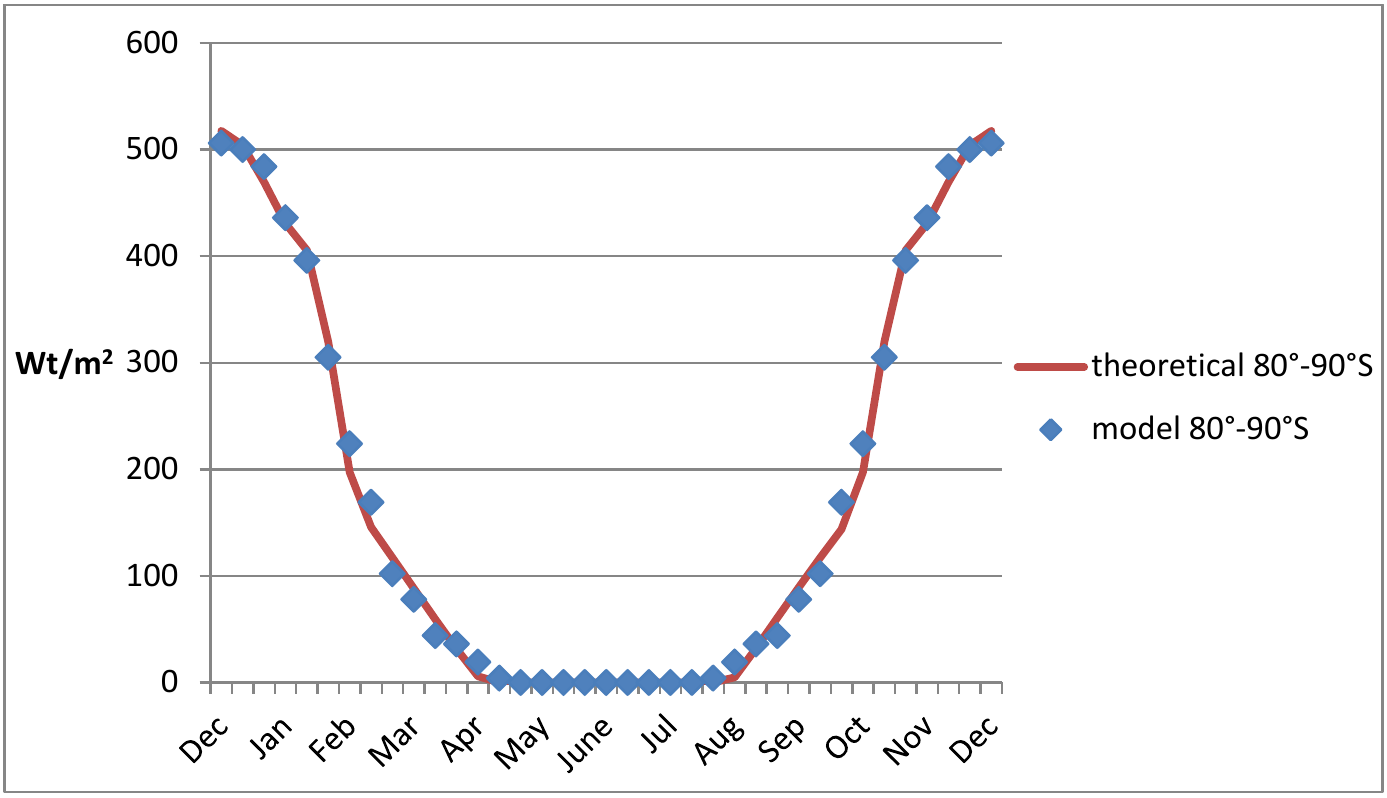}}}		
		\caption{Curve fitting for $0^\circ-10^\circ$, $40^\circ-50^\circ$ and $80^\circ-90^\circ$ latitudinal belts in the Northern and Southern Hemisphere.}
		\label{fig:3}
	\end{center}
\end{figure}

The resulting curves for the $0^\circ-10^\circ$, $40^\circ-50^\circ$ and $80^\circ-90^\circ$ latitudinal belts in the Northern and Southern Hemispheres are presented in Figure \ref{fig:3}. In each figure the blue curve corresponds to the data obtained from the model, and the red one is the approximation curve.

In order to verify how well the values from the proposed model are replicated by the approximation, the $R$ squared test was applied. The $R^2$ coefficient was computed using \textquotedblleft Data Analysis\textquotedblright \space add-in in MS Excel. The \textquotedblleft Regression\textquotedblright \space analysis tool was used. The results obtained are shown in Table \ref{table1}.

\begin{table}[ht]
	
	\resizebox{0.5\textwidth}{!} {	
		\begin{tabular}{cccc}\hline
			Latitudinal belt	&$R^2$ & Latitudinal belt &$R^2$  \\ \hline
			$0^\circ-10^\circ$N	& 0.98  &$0^\circ-10^\circ$S  & 0.96 \\ 
			$10^\circ-20^\circ$N& 0.96 &$10^\circ-20^\circ$S  & 0.95 \\ 
			$20^\circ-30^\circ$N& 0.99 &$20^\circ-30^\circ$S  & 0.99 \\ 
			$30^\circ-40^\circ$N& 0.99 &$30^\circ-40^\circ$S  & 0.99 \\ 
			$40^\circ-50^\circ$N& 0.99 &$40^\circ-50^\circ$S  & 0.99 \\ 
			$50^\circ-60^\circ$N& 0.99  &$50^\circ-60^\circ$S  & 0.99 \\ 
			$60^\circ-70^\circ$N& 0.99 &$60^\circ-70^\circ$S  & 0.99 \\ 
			$70^\circ-80^\circ$N& 0.99 &$70^\circ-80^\circ$S  & 0.99 \\
			$80^\circ-90^\circ$N& 0.99 &$80^\circ-90^\circ$S & 0.99
			\\\hline 
		\end{tabular}  }
		\caption{The value of $R^2$ coefficient for each latitudinal belt in the Northern and Southern Hemisphere.}	
		\label{table1}
	\end{table}
	
	The $R^2$ coefficient obtained for all the latitudinal belts is greater than the required 95\% confidence level. In particular, for the latitudinal belts beyond the $20^\circ$, its value is 99\%. Thus the conclusion can be drawn that the fitted curves are appropriate for approximating the values obtained from the proposed model. 
	
	\section{Simulating the seasonal variations of insolation within the C-GOLDSTEIN model}
	The calculations of insolation were initially performed in one of the subroutines of the global climate model where the atmosphere is initialized prior to the start of iterations. Firstly, we have replaced the average annual values used there by the annual average values obtained from our proposed model \cite{prakhova2014new}. Note that C-GOLDSTEIN has a latitudinal resolution of $20^\circ$ near the polar regions. Thus, we chose to extend the value for the $70^\circ-80^\circ$ belt to the $80^\circ-90^\circ$ latitudinal belt. 
	
	Both sets of the yearly averages were compared with satellite data from the NASA Langley Research Centre Atmospheric Science Data Centre Surface meteorological and Solar Energy (SSE) web portal supported by the NASA LaRC POWER Project.\footnote{\url{https://eosweb.larc.nasa.gov/sse/global/text/22yr_toa_dwn}}
	Note that the data in the source is given in terms of $1^\circ$ resolution so we have averaged this over the $10^\circ$ latitudinal belt. The comparison of the results is shown in Table \ref{table2}. 
	
	\newcolumntype{L}[1]{>{\raggedright\let\newline\\\arraybackslash\hspace{0pt}}m{#1}}
	\newcolumntype{C}[1]{>{\centering\let\newline\\\arraybackslash\hspace{0pt}}m{#1}}
	\newcolumntype{R}[1]{>{\raggedleft\let\newline\\\arraybackslash\hspace{0pt}}m{#1}}
	
	\begin{table}[ht]
		
		\resizebox{1\textwidth}{!} {	
			\begin{tabular}{C{2.5cm}C{2.5cm}C{2.5cm}C{2.5cm}C{2.5cm}C{2.5cm}}\hline
				Latitudinal belt & Satellite data (Wt/m$^2$) & Proposed insolation model (Wt/m$^2$) &   The insolation component of C-GOLDSTEIN (Wt/m$^2$) & \multicolumn{2}{c}{Accuracy} \\\cline{5-6}
				&  &  &  & Proposed insolation model & The insolation component of C-GOLDSTEIN \\\hline
				$0^\circ-10^\circ$ & 415.00 & 408.40 & 420.67 & 0.99 & 0.99\\
				$10^\circ-20^\circ$ & 398.45 & 381.65 & 407.33 & 0.97 & 0.98\\
				$20^\circ-30^\circ$ & 378.29 & 376.20 & 380.33 & 0.99 & 0.99\\
				$30^\circ-40^\circ$ & 359.76 & 346.50 & 339.00 & 0.99 & 0.94\\
				$40^\circ-50^\circ$ & 304.33 & 300.83 & 295.00 & 0.99 & 0.97\\
				$50^\circ-60^\circ$ & 257.78 & 251.33 & 253.00 & 0.97 & 0.98\\
				$60^\circ-70^\circ$ & 220.00 & 213.22 & 217.00 & 0.97 & 0.99\\
				$70^\circ-80^\circ$ & 182.02 & 172.50 & 192.00 & 0.95 & 0.95\\
				$80^\circ-90^\circ$ & 169.89 & 172.50 & 192.00 & 0.98 & 0.87\\ \hline
				Average & & & & 0.98 & 0.96\\ \hline
			\end{tabular} }
			\caption{The comparison of the results of initial and proposed insolation models.}	
			\label{table2}
		\end{table}
		
		The results obtained indicate a 2\% increase in the average accuracy compared to the insolation values used previously in C-GOLDSTEIN. Also, there is a significant increase in accuracy for the furthest polar belt.
		
		After the comparison we reinitialised the atmosphere starting from zero initial conditions and ran C-GOLDSTEIN in SPINUP mode leaving all other parameters set to zero (such as carbon dioxide growth rate etc.) in order to obtain suitable initial conditions. The model was run remotely in high performance computing environment.
		
		The curves obtained in Section \ref{Curve} have then been incorporated into the main loop of the C-GOLDSTEIN simulation model. In this way, the insolation computations are performed at each time step. The model was then run with the modified code using the results obtained after the SPINUP run as the initial conditions. The time step was reduced to 1 day for the ocean (compared to the initial 1.46 days). The ocean-atmosphere time step ratio was kept unchanged (the atmospheric time step is half of the ocean one). Note that a 360-day calendar was used for simplicity (i.e. each month has 30 days). 
		
\begin{figure}
	\begin{center}
		\subfigure[December.]{
			\resizebox*{3.7cm}{!}{\includegraphics{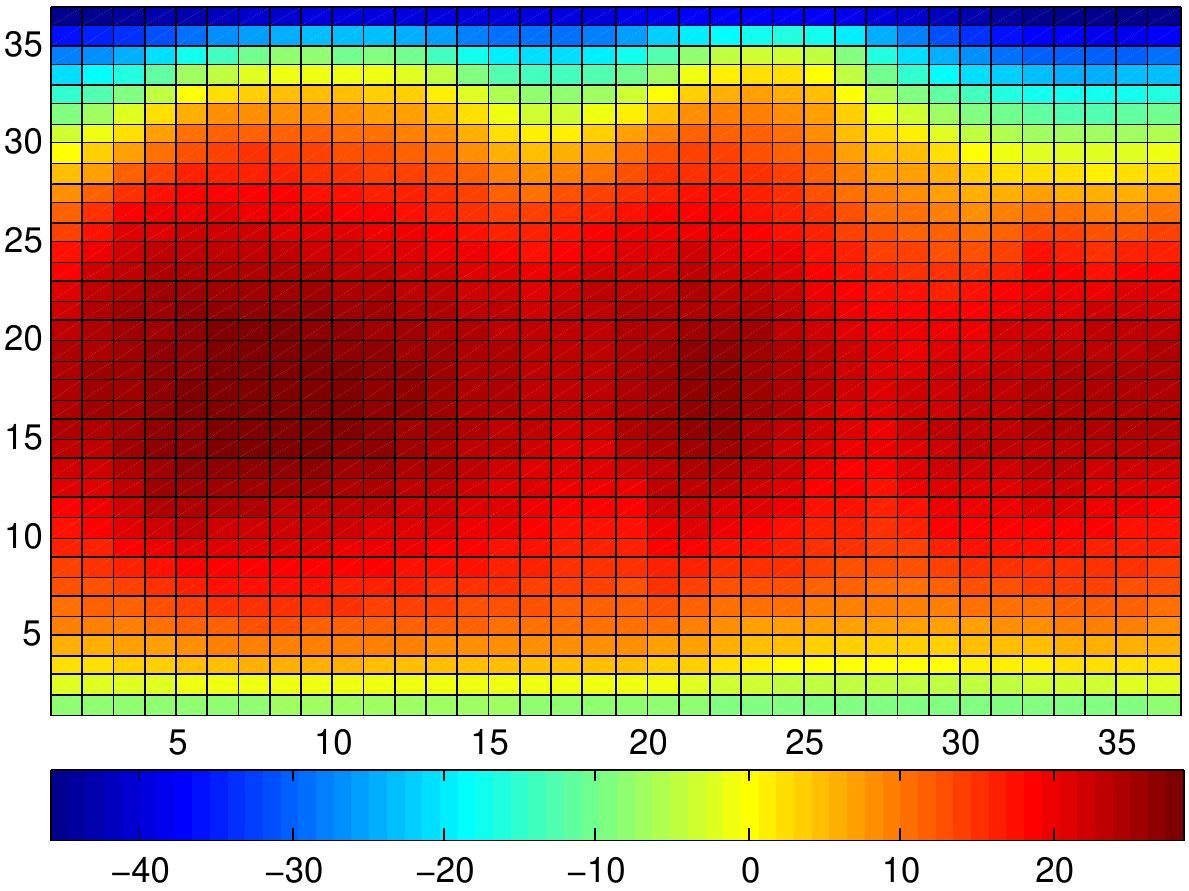}}}	
		\subfigure[January.]{
			\resizebox*{3.7cm}{!}{\includegraphics{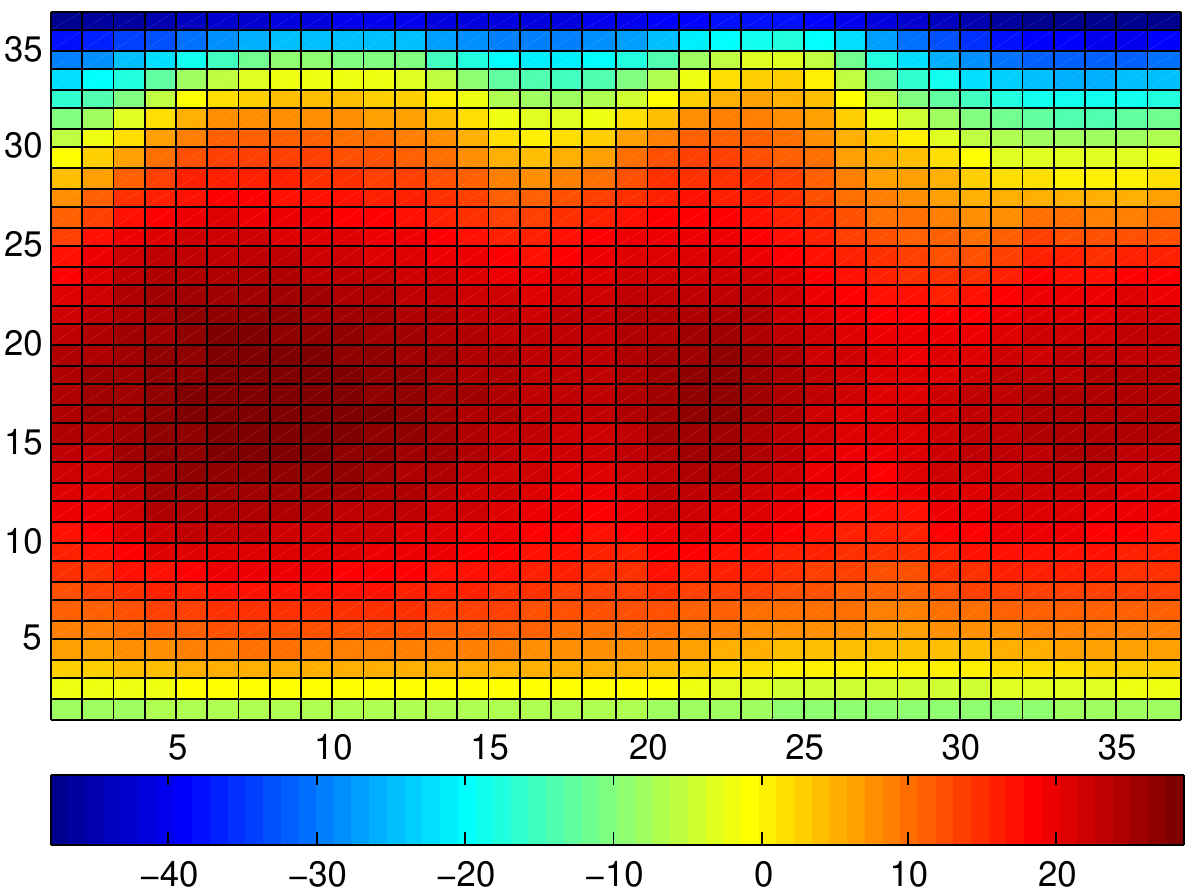}}}
		\subfigure[February.]{
			\resizebox*{3.7cm}{!}{\includegraphics{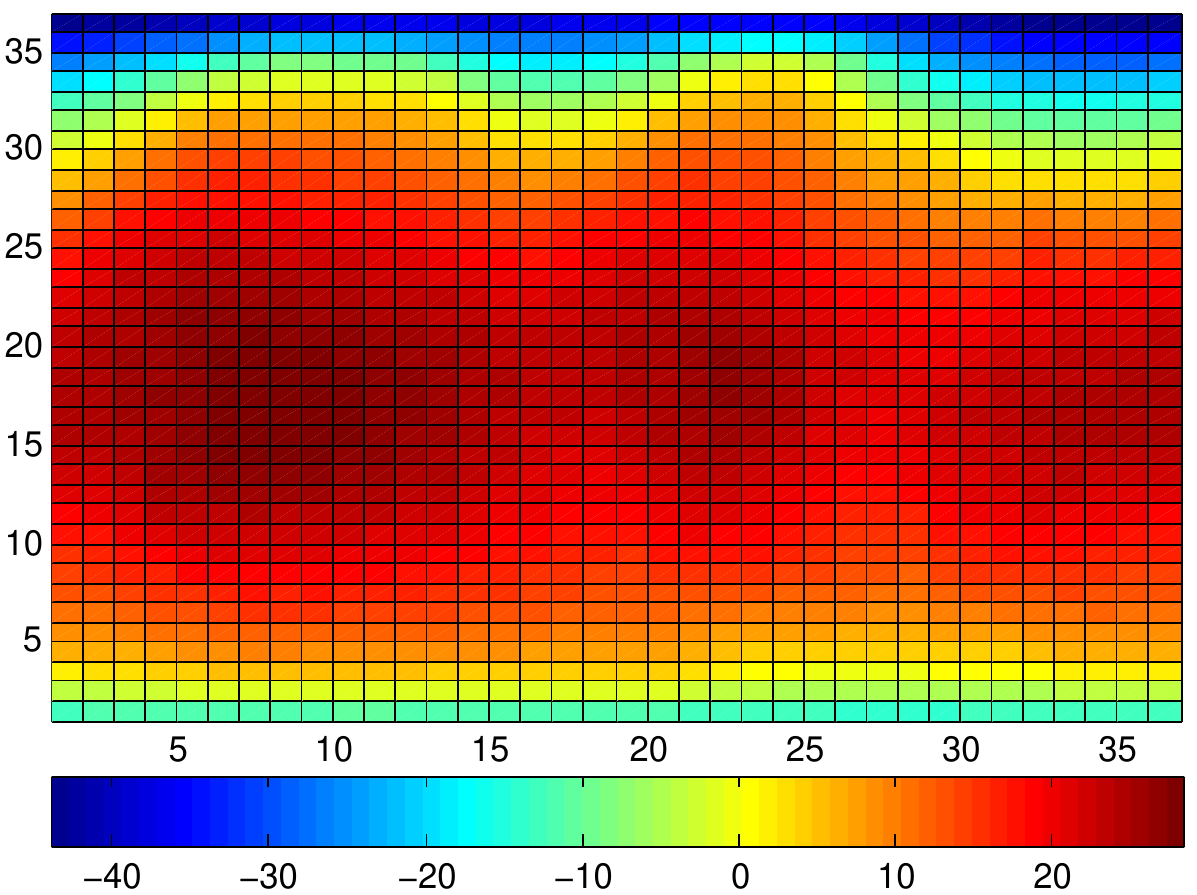}}}
		
		\subfigure[March.]{
			\resizebox*{3.7cm}{!}{\includegraphics{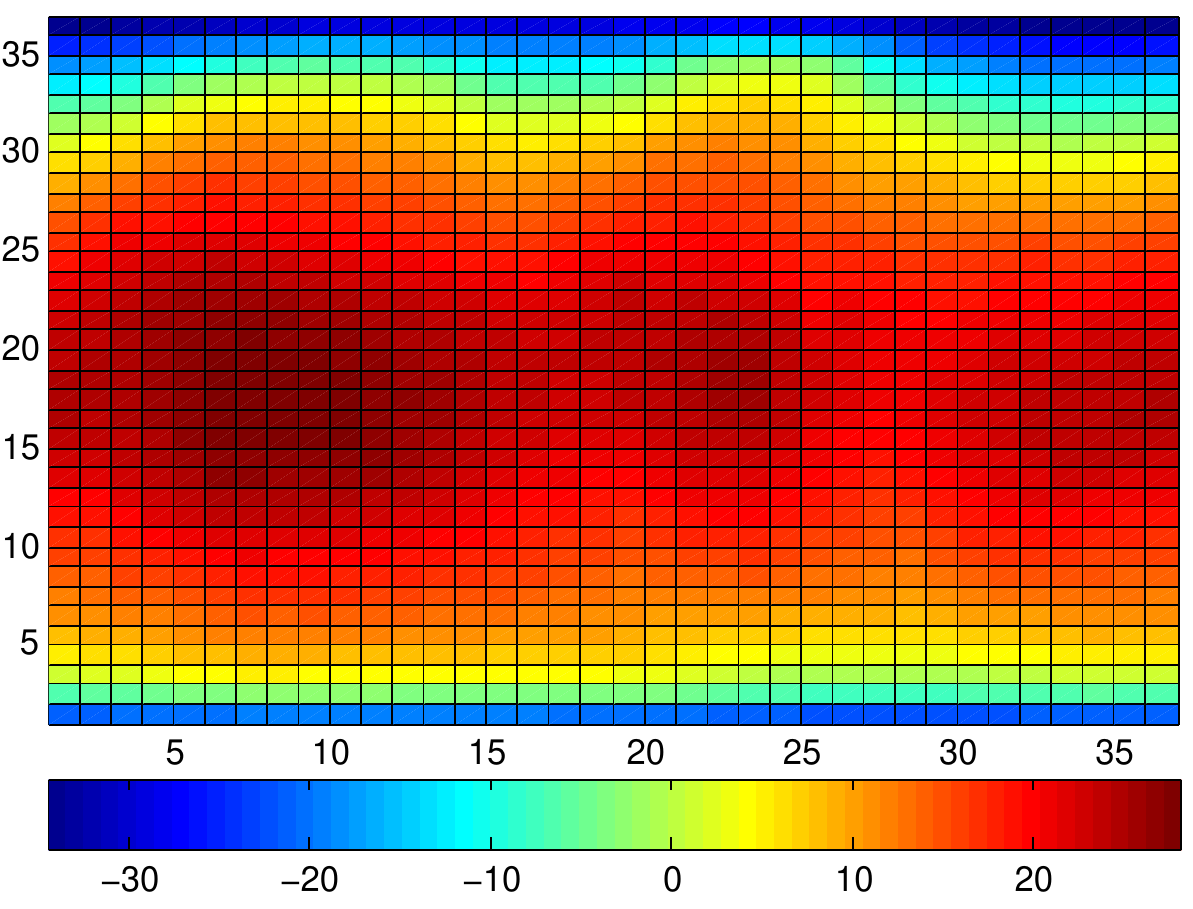}}}	
		\subfigure[April.]{
			\resizebox*{3.7cm}{!}{\includegraphics{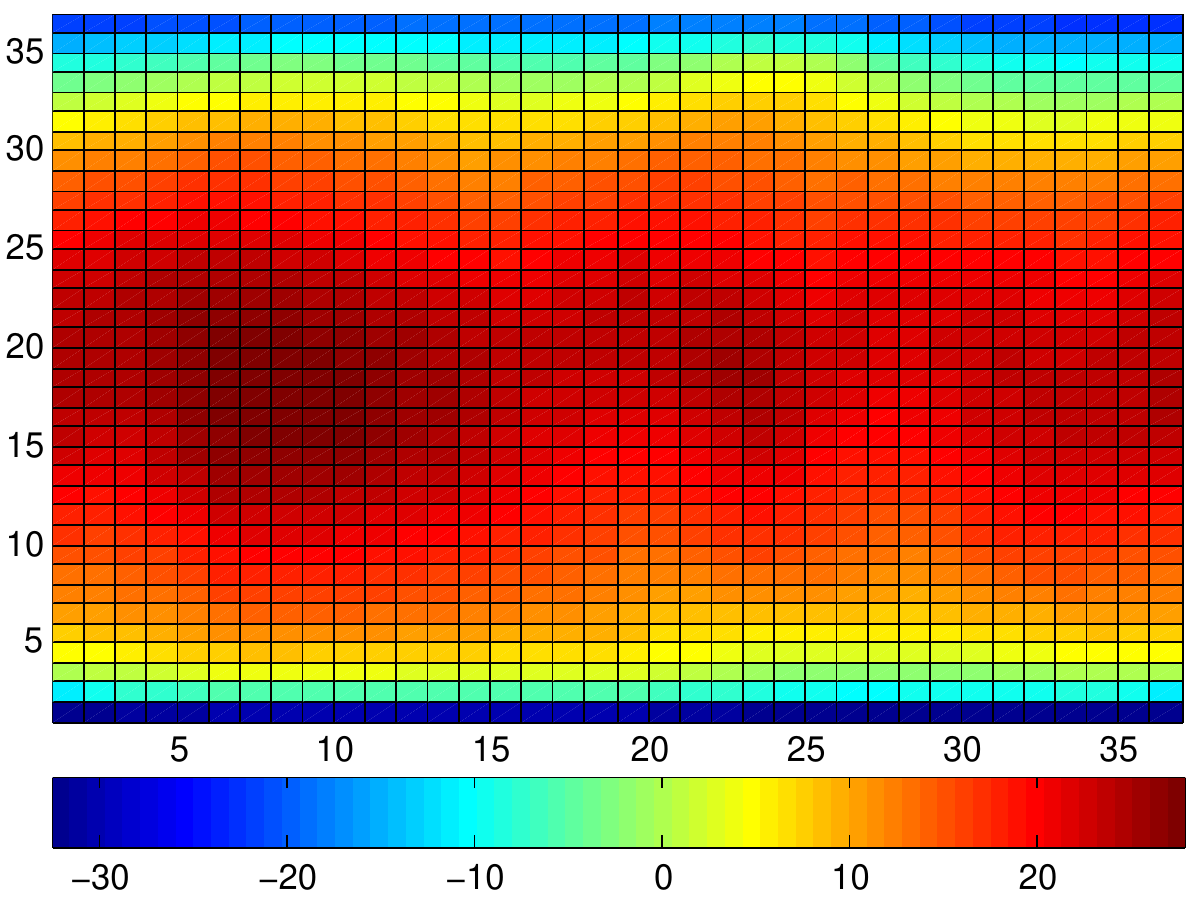}}}
		\subfigure[May.]{
			\resizebox*{3.7cm}{!}{\includegraphics{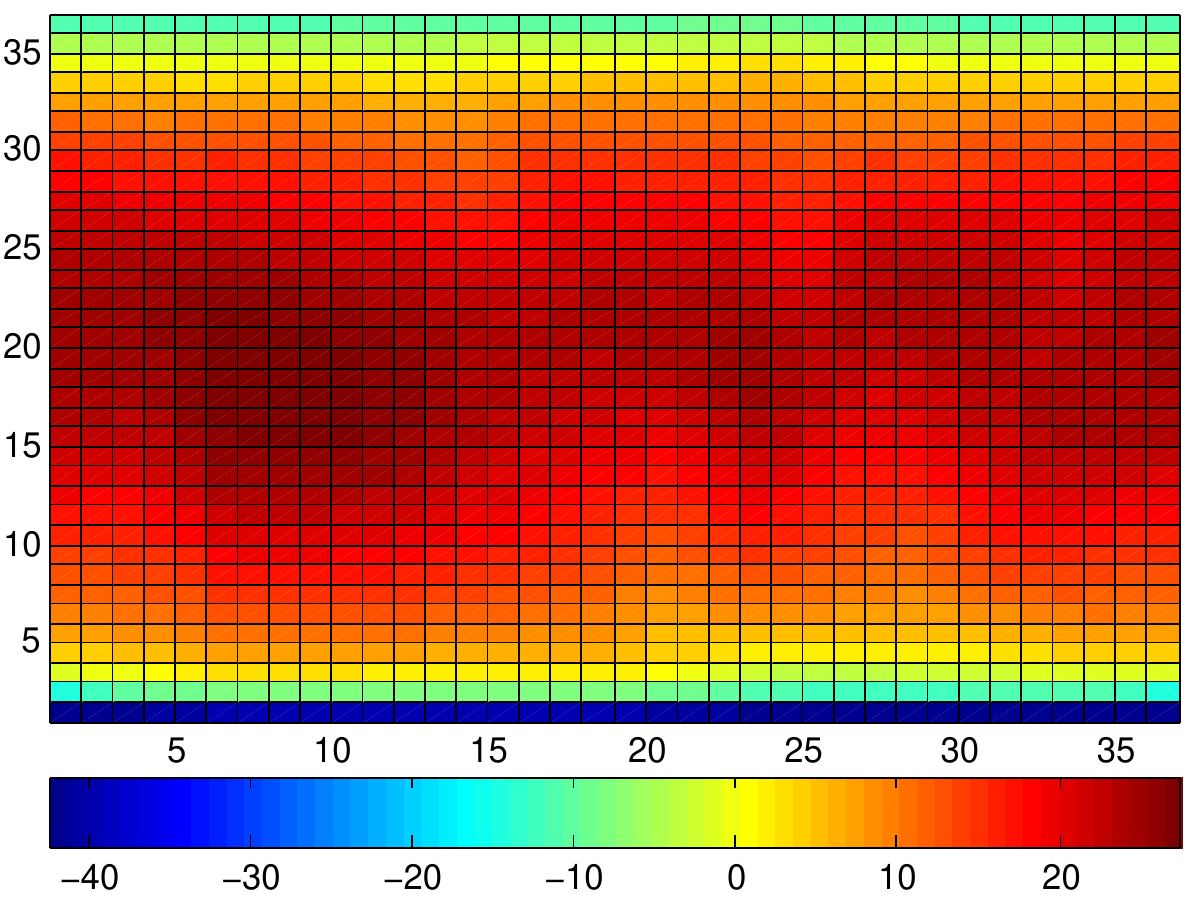}}}
		
		\subfigure[June.]{
			\resizebox*{3.7cm}{!}{\includegraphics{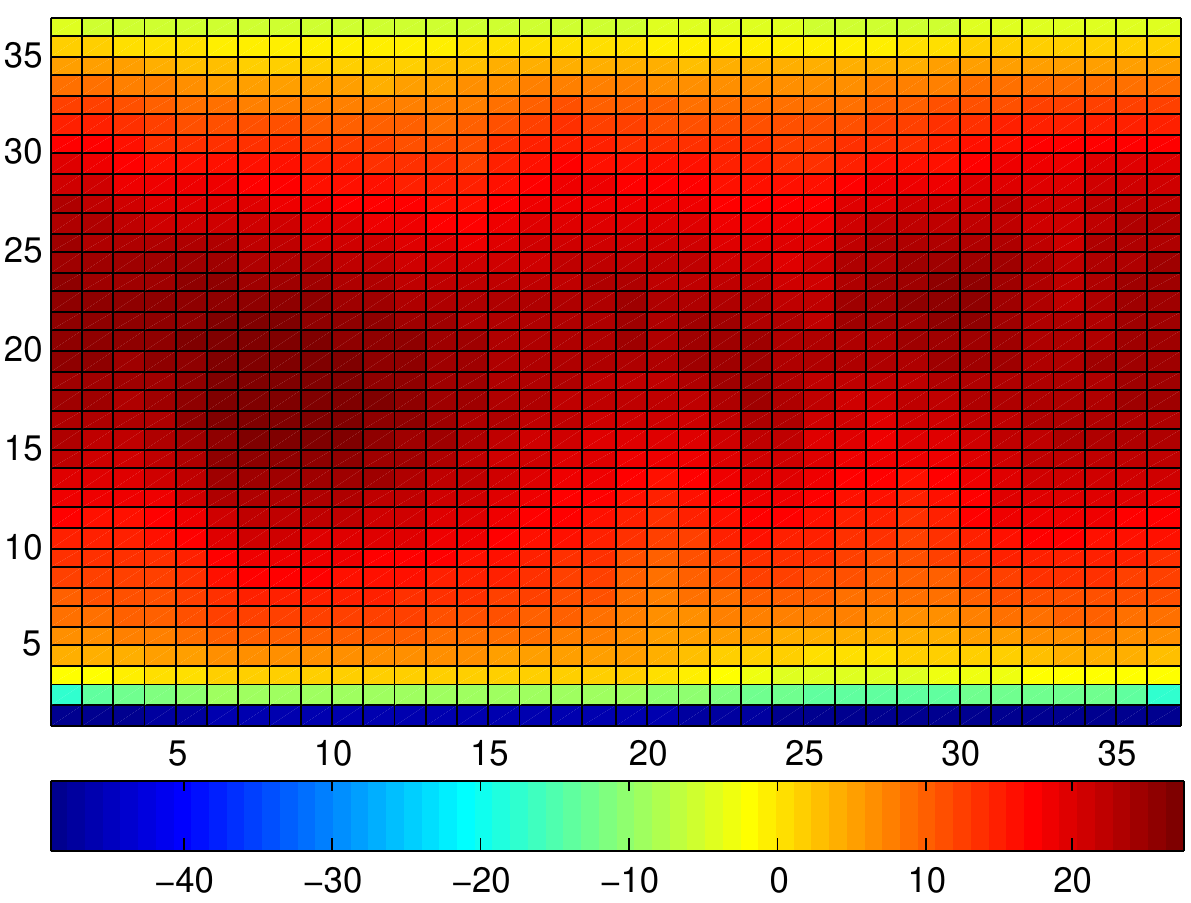}}}	
		\subfigure[July.]{
			\resizebox*{3.7cm}{!}{\includegraphics{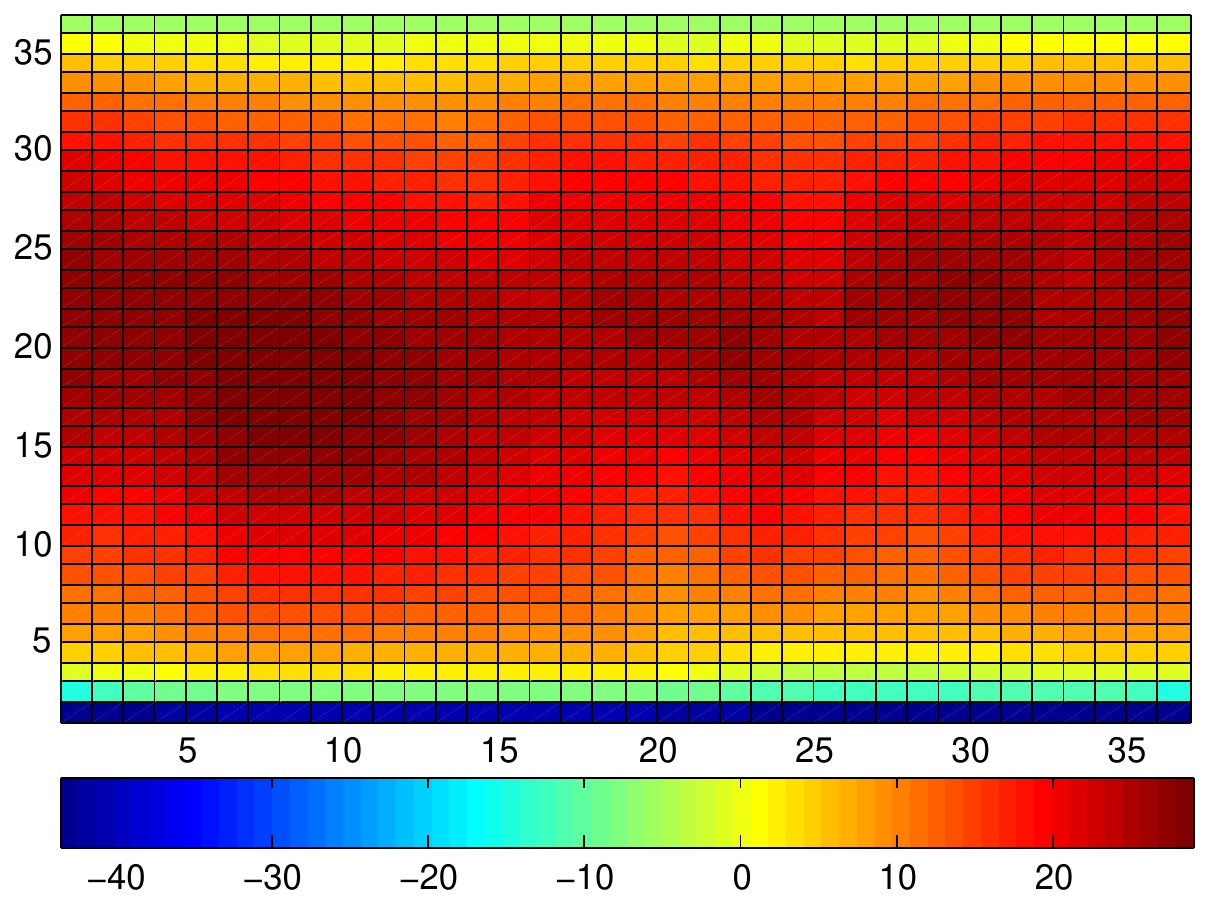}}}	
		\subfigure[August.]{
			\resizebox*{3.7cm}{!}{\includegraphics{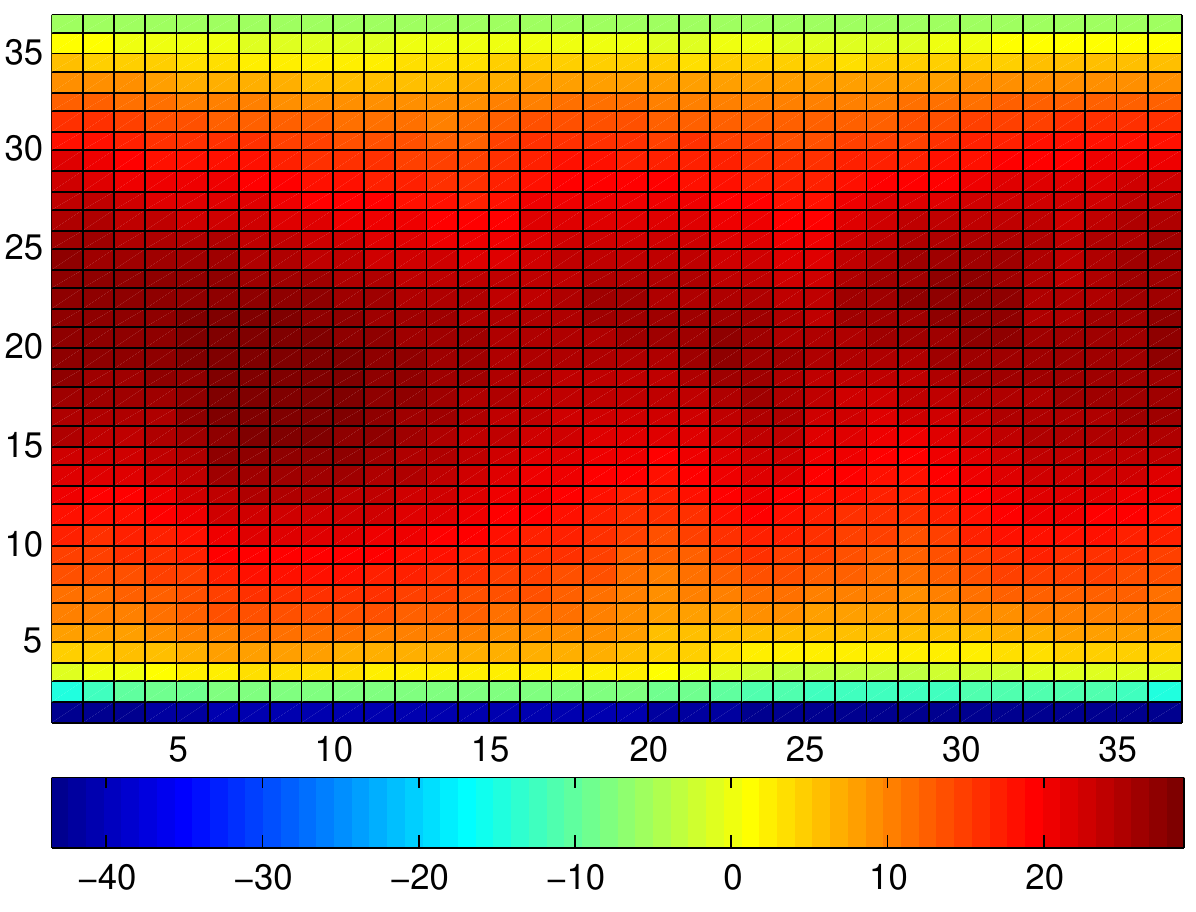}}}
		
		\subfigure[September.]{
			\resizebox*{3.7cm}{!}{\includegraphics{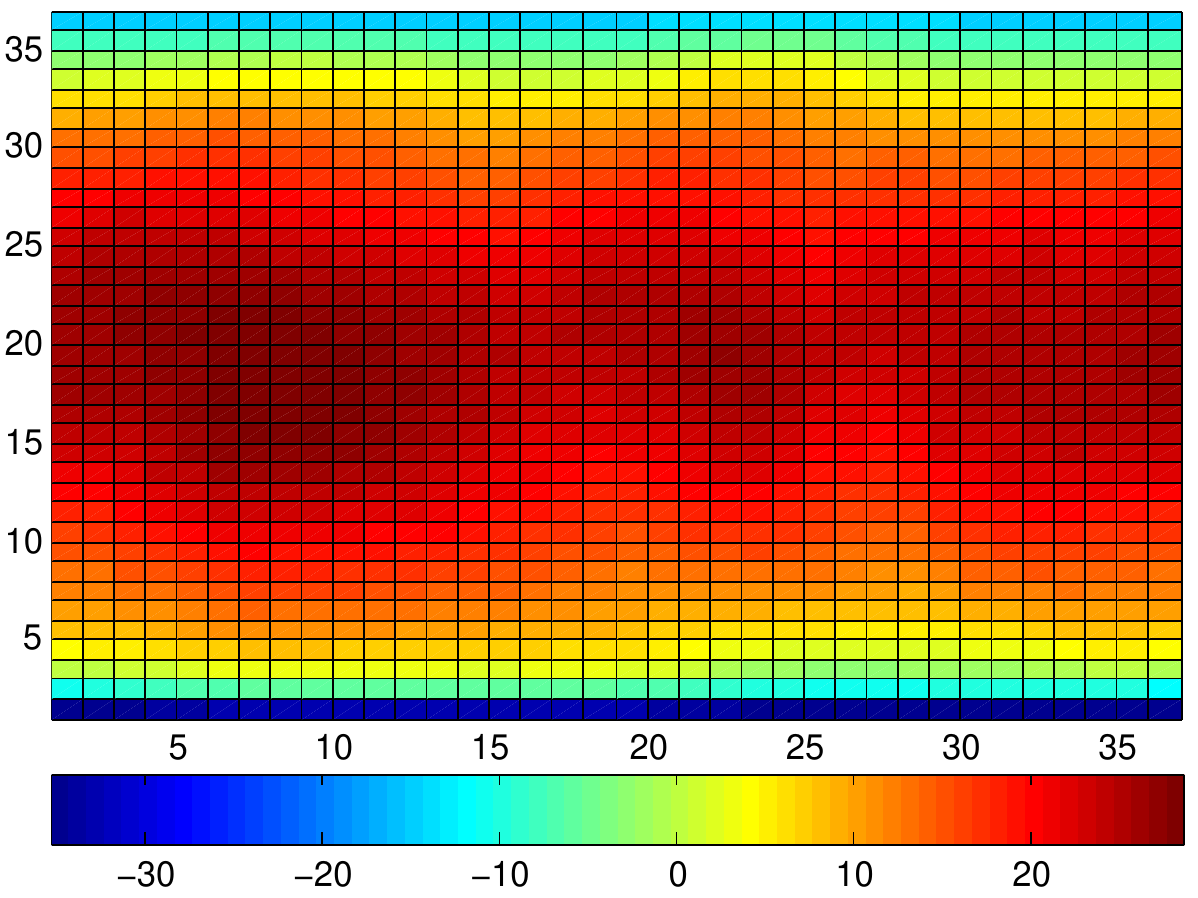}}}	
		\subfigure[October.]{
			\resizebox*{3.7cm}{!}{\includegraphics{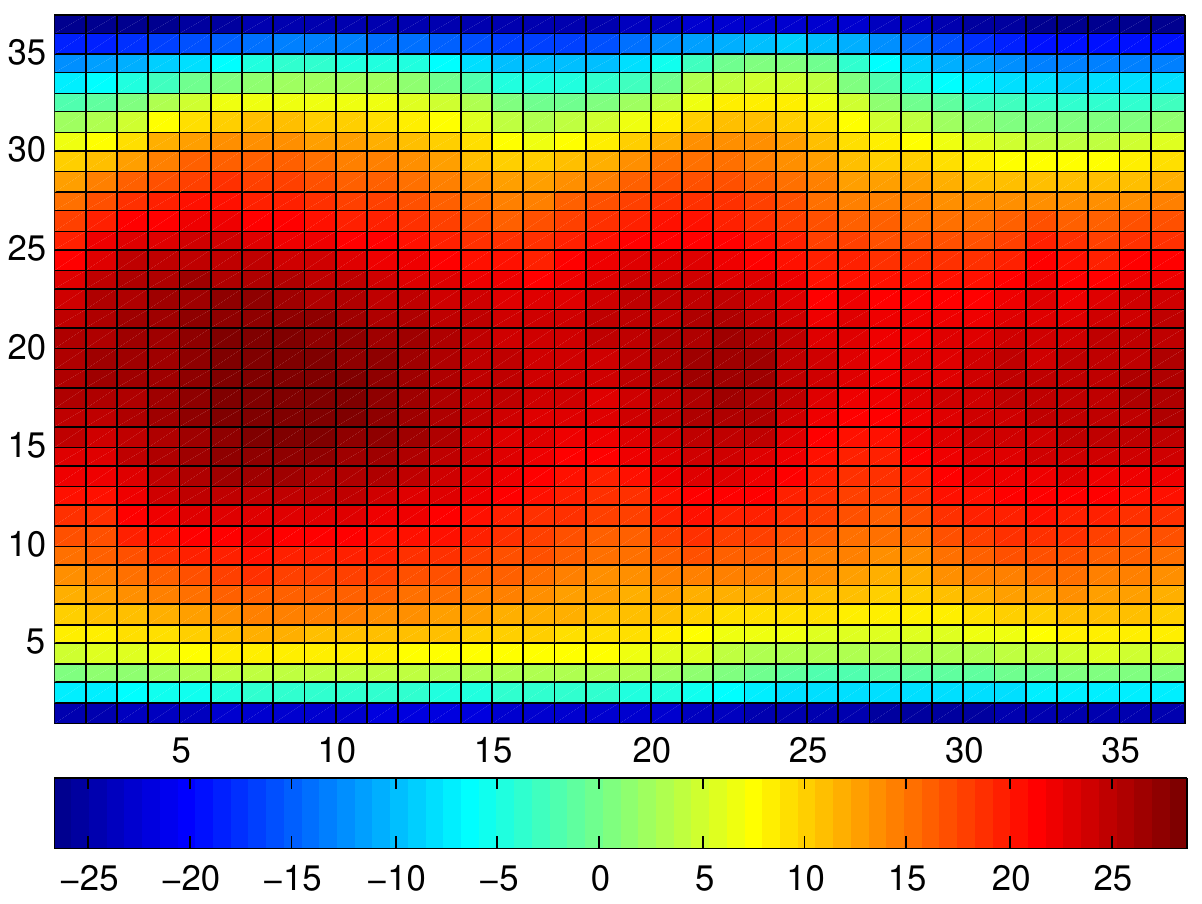}}}	
		\subfigure[November.]{
			\resizebox*{3.7cm}{!}{\includegraphics{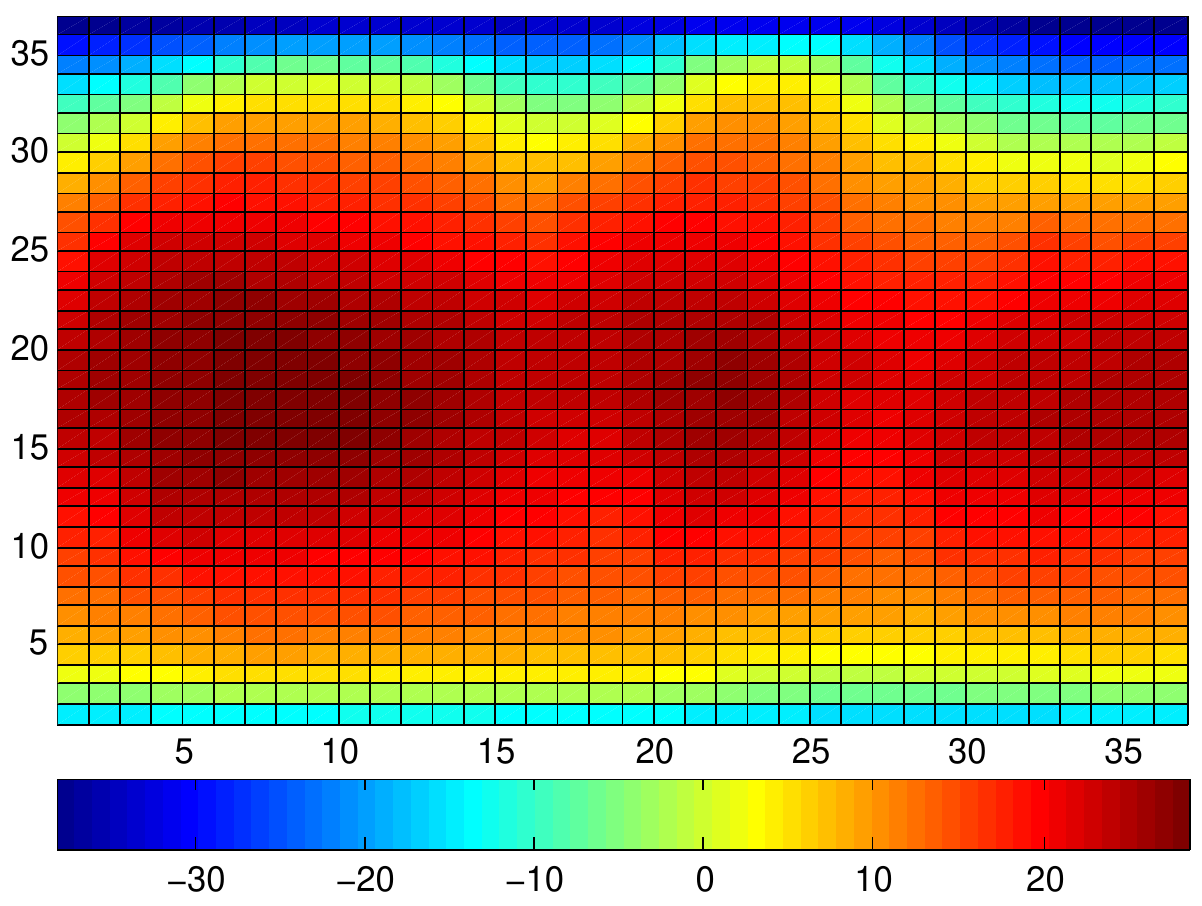}}}	
		\caption{Temperature distribution throughout the year.}
		\label{fig:4}
	\end{center}
\end{figure}

		The results for the 21st day of each month are illustrated in Figure \ref{fig:4}. The figures were obtained with a small modification to the MATLAB plotting subroutine provided together with the model software.
		
		The results can be compared with the monthly temperature distribution maps from 
		National Centres for Environmental Predictions (NCEP)/National Centre for Atmospheri Research (NCAR) Reanalysis Project.\footnote{\url{ http://geog.uoregon.edu/envchange/clim_animations/flash/tmp2m.html}}
		
		Clearly, the obtained temperature distributions are realistic and follow all the main patterns in the actual temperature distributions from NCEP/NCAR, such as the maintenance of hot temperature throughout the year for the equatorial regions, the rotation of the winter and summer seasons for the Northern and Southern Hemispheres, extreme low observed temperatures for the polar regions during their winter seasons, and distinct temperature variations due to the location of continents.
		
		\section{Conclusion}
		In this paper, we have incorporated the seasonal variations of insolation into the global climate model C-GOLDSTEIN. Firstly, the latitudinal curves of insolation throughout the year obtained from the authors\textquoteright \space earlier work were approximated by the functions of a more suitable form for computation. These curves were then incorporated into the main loop of C-GOLDSTEIN. The model was then run remotely in high performance computing environment. 
		
		Realistic monthly temperature distributions have been obtained after running the global climate model with the new insolation component. Also, the average accuracy of modelling the insolation within C-GOLDSTEIN has been increased from 96\% to 98\%.  In addition, new types of experiments can now be performed with the C-GOLDSTEIN model, because the calculations can now be performed for any particular time of the year. For example, consequences of random temporal variations of insolation on temperature can now be examined.

\newpage

  \bibliographystyle{elsarticle-num}
\bibliography{mybiblio}

\end{document}